\documentclass[twocolumn,amsmath,amssymb,aps,pre,longbibliography]{revtex4-1}
\usepackage{graphicx}% Include figure files
\usepackage{bm}% bold math

\usepackage{amsfonts}
\usepackage{amsmath,amsfonts,amsthm,bm}

\usepackage[utf8]{inputenc}

\begin{document}
\title{Three-dimensional vortex dipole solitons in self-gravitating systems}

%\author{V. M. Lashkin}
%\email{vlashkin62@gmail.com} \affiliation{$^1$Institute for
%Nuclear Research, Pr. Nauki 47, Kyiv 03028, Ukraine}
%\affiliation{$^2$Space Research Institute, Pr. Glushkova 40 k.4/1,
%Kyiv 03187,  Ukraine}

%\author{Volodymyr M. Lashkin$^{1,2}$,
%Zahida Ehsan$^{3,4}$  and Nazia Batool$^{4}$}
% \affiliation{$^{1}$ Institute for
%Nuclear Research, Pr. Nauki 47, Kyiv 03028, Ukraine}
%\affiliation{$^{2}$Space Research Institute, Pr. Glushkova 40
%k.4/1, Kyiv 03187, Ukraine} \affiliation{$^{3}$ SPAR and The
%Landau-Feynman Laboratory for Theoretical Physics, Department of
%Physics, CUI, Lahore Campus 54000, Pakistan } \affiliation{$^{4}$
%National Centre for Physics, Shahdara Valley Road, Islamabad
%45320, Pakistan.} \email{vlashkin62@gmail.com}

%\author{Volodymyr M. Lashkin$^{1,2}$}
%\email{vlashkin62@gmail.com} \affiliation{$^{1}$ Institute for
%Nuclear Research, Pr. Nauki 47, Kyiv 03028, Ukraine}
%\affiliation{$^{2}$Space Research Institute, Pr. Glushkova 40
%k.4/1, Kyiv 03187, Ukraine}
% \author{Zahida Ehsan$^{3,4}$}
% \affiliation{$^{3}$ SPAR and The
%Landau-Feynman Laboratory for Theoretical Physics, Department of
%Physics, CUI, Lahore Campus 54000, Pakistan }
%\author{Nazia Batool$^{4}$}  \affiliation{$^{4}$
%National Centre for Physics, Shahdara Valley Road, Islamabad
%45320, Pakistan}

\author{Volodymyr M. Lashkin}
\email{vlashkin62@gmail.com} \affiliation{Institute for Nuclear
Research, Pr. Nauki 47, Kyiv 03028, Ukraine} \affiliation{Space
Research Institute, Pr. Glushkova 40 k.4/1, Kyiv 03187, Ukraine}
\author{Oleg K. Cheremnykh}
\affiliation{Space Research Institute, Pr. Glushkova 40 k.4/1,
Kyiv 03187, Ukraine}
\author{Zahida Ehsan}
\affiliation{SPAR and The Landau-Feynman Laboratory for
Theoretical Physics, Department of Physics, CUI, Lahore Campus
54000, Pakistan } \affiliation{ National Centre for Physics,
Shahdara Valley Road, Islamabad 45320, Pakistan}
\author{Nazia Batool}
\affiliation{ National Centre for Physics, Shahdara Valley Road,
Islamabad 45320, Pakistan}

\begin{abstract}
We derive the nonlinear equations governing the dynamics of
three-dimensional (3D) disturbances in a nonuniform rotating
self-gravitating fluid under the assumption that the
characteristic frequencies of disturbances are small compared to
the rotation frequency. Analytical solutions of these equations
are found in the form of the 3D vortex dipole solitons. The method
for obtaining these solutions is based on the well-known
Larichev-Reznik procedure for finding two-dimensional nonlinear
dipole vortex solutions in the physics of atmospheres of rotating
planets. In addition to the basic 3D $x$-antisymmetric part
(carrier), the solution may also contain radially symmetric
(monopole) or/and antisymmetric along the rotation axis ($z$-axis)
parts with arbitrary amplitudes, but these superimposed parts
cannot exist without the basic part. The 3D vortex soliton without
the superimposed parts is extremely stable. It moves without
distortion and retains its shape even in the presence of an
initial noise disturbance. The solitons with parts that are
radially symmetric or/and $z$-antisymmetric turn out to be
unstable, although at sufficiently small amplitudes of these
superimposed parts, the soliton retains its shape for a very long
time.
\end{abstract}

\maketitle

\section{Introduction}

The study of the dynamics of self-gravitating systems began with
the pioneering work by Jeans \cite{Jeans1929}, who showed that,
within the framework of the Euler equations with a gravitational
potential obeying the Poisson equation, perturbations with the
wavelength $\lambda$ greater than the Jeans wavelength
$\lambda_{J}$ are unstable. It is believed that the Jeans
instability is the source of the emergence of structures in the
Universe and this problem remains one of the most important in
astrophysics. Even Jeans himself put forward the hypothesis that
stars, star clusters, and galaxies  arose as a result of this
instability -- a process resembling condensation in an ordinary
imperfect gas. Subsequently, linear instability in a
self-gravitating system and the Jeans criterion were investigated
by many authors, starting with Fermi and Chandrasekhar
\cite{Fermi1953,Chandrasekhar_book}, for cases of system rotation,
inhomogeneity of the equilibrium density, the presence of magnetic
fields and dust, accounting for kinetic effects, dissipation, etc.
\cite{KineticJeans2004,Ehsan2007,Ehsan2008,Maklund2008,Kremer2018}.
The linear instability theory is only valid when the amplitude of
the perturbations is so small that the nonlinearity can be
neglected. A direct demonstration of the possibility of the
emergence of structures from spontaneously arising fluctuations at
the nonlinear stage of instability (i.e., nonlinear evolution) is
an extremely difficult task. In turn, coherent nonlinear
structures themselves, such as solitons and vortices, have been
studied for a long time
\cite{Petviashvili_book1992,Kivshar_book2003,Horton1996,Manton2004,Dauxois2006}.
In a broad sense, a soliton is a localized structure (not
necessarily one-dimensional) resulting from the balance of
dispersion and nonlinearity effects. Two- or three-dimensional
solitons with embedding vorticity are usually called vortex
solitons. Note that below we sometimes refer to vortex solitons
simply as vortices, although vortices are usually understood as
structures (for example, vortices in hydrodynamics) in media
without dispersion or where the role of dispersion does not
matter. One-dimensional solitons are usually stable, while
multidimensional solitons often turn out to be unstable and the
most well-known phenomena in this case are wave collapse and wave
breaking \cite{Berge1998,Zakharov_UFN2012}. Nevertheless, there
are many examples of stable multidimensional solitons. The reason
is usually the specific nature of the nonlinearity (nonlocal
\cite{Lashkin2006,Lashkin2007PLA}, saturable
\cite{Laedke1984,Lashkin2020} or additional higher order
nonlinearity) and dispersion \cite{Kivshar_book2003}. In some
cases, a specific form of nonlinearity (the Poisson bracket
nonlinearity) leads to the presence of an infinite number of
integrals of motion (Casimir invariants), which causes the
stability of the corresponding multidimensional solitons
\cite{Makino1981,Williams1982,Lashkin2017,Petviashvili_book1992}.
Multidimensional solitons have also been intensively studied in
scalar models of quantum field theory
\cite{Friedberg1976,Lee1992}(see, e. g., recent paper
\cite{Morris2021}). The stability of such solitons follows from
the well-known Derrick criterion \cite{Derrick1964}.

Solitons in self-gravitating systems were apparently first
considered in Ref.~\cite{Mikhailovskii1977}, where the Jeans
perturbations of finite amplitude were studied and it was shown
that they can propagate in the form of envelope solitons. Later
on, solitons in self-gravitating systems were studied in a number
of works. Nonlinear waves in a self-gravitating isothermal fluid
were considered in Ref.~\cite{Yueh1981}. Within the framework of
the same model, the nonlinear Schr\"{o}dinger equation (NLS)
\cite{Ono1994, Zhang1995,Zhang1998} and the sine-Gordon equation
\cite{Gotz1988} were derived and their soliton solutions were
presented. One-dimensional nonlinear waves and solitons in
self-gravitating fluid systems, with a particular emphasis on
applications to molecular clouds, were studied in
\cite{Adams1994}. Self-gravitating fluid dynamics and
instabilities along with solitons were discussed in
Ref.~\cite{Semelin2001}. Solitons in self-gravitating dusty
plasmas were considered on the basis of the extended Korteveg-de
Vries (KdV) equation in Ref.~\cite{Verheest1997}, and Alfv\'{e}n
ordinary, cusp solitons and modulational instability in a
self-gravitating magneto-radiative plasma were studied in
Ref.~\cite{Masood2010}. Solitary waves in self-gravitating
molecular clouds were investigated in Ref.~\cite{Verheest2005}. In
the above works, only one-dimensional solitons were considered.

Here we would like to note that in this paper we are interested in
nonlinear structures in self-gravitating systems exclusively
within the framework of the classical fluid model, since in recent
years solitons and vortex solitons in self-gravitating
Bose-Einstein condensates (BEC) (nonlinear matter waves) based on
the Gross-Pitaevskii equation for quantum mechanical wave function
have been intensively studied \cite{Yakimenko2021}.

We are interested in rotating self-gravitating systems and
nonlinear perturbations with characteristic frequencies that are
much lower than the rotational frequency of the system. Under such
assumptions, two-dimensional (2D) nonlinear structures in
self-gravitating systems were first considered in
Ref.~\cite{Fridman1991}, where the corresponding nonlinear
equation was derived, which coincides with the well-known Charney
equation in geophysics \cite{Charney1948} and describing nonlinear
Rossby waves in atmospheres of rotating planets and in oceans (in
plasma physics, this equation is known as the Hasegawa-Mima
equation \cite{Hasegawa1978}, and the rotation frequency is
replaced by the gyrofrequency in an external magnetic field). In
what follows, we refer to this equation as the 2D
Charney-Hasegawa-Mima (CHM) equation. The analytical solution to
this equation is the 2D dipole solitary vortex obtained for the
first time in Refs.~\cite{Larichev1976a,Larichev1976b} and known
as the Larichev-Reznik dipole vortex (sometimes called a modon).
Subsequently, this solution and some of its generalizations in the
form of dipole vortices were used in many areas of nonlinear
geophysics, as well as for describing nonlinear drift waves in
plasmas \cite{Flierl1987,Petviashvili_book1992,Stenflo2009}. In
rotating self-gravitating systems, dipole vortex solutions were
obtained for magnetized plasmas \cite{Jovanovich1990} and bounded
systems (so called global vortices) \cite{Shukla1993}. Regular
structures in a rotating dusty self-gravitating fluid system were
also studied in Ref.~\cite{Zinzadze2000}. Some generalizations of
these structures, including monopole vortices, were studied in
Refs.~\cite{Abrahamyan2016,Abrahamyan2020}. Nonlinearly coupled
Rossby-type and inertio-gravity waves in self-gravitating systems
were considered in Ref.~\cite{Pokhotelov1998}, nonlinear vortex
chains in Ref.~ \cite{Shukla1995}. The emergence of vortices in
self-gravitating gaseous discs was demonstrated by numerical
simulation in Ref.~\cite{Rice2009}.

To avoid misunderstandings with terminology, it should be noted
that, generally speaking, there are two types of vortex solitons.
The Larichev-Reznik soliton (as well as the solitons considered in
the presented paper) arises in models with linear dispersion
$\omega_{\mathbf{k}}$ ($\omega$ and $\mathbf{k}$ are the frequency
and wave vector,respectively) of the acoustic type
($\omega_{\mathbf{k}} \rightarrow 0$ as $\mathbf{k}\rightarrow 0$)
and is a dipole vortex soliton representing a cyclone-anticyclone
dipole pair that rotate in opposite directions. In models with
linear dispersion of the optical type
($\omega_{\mathbf{k}}\rightarrow \omega_{c}$ as
$\mathbf{k}\rightarrow 0$, where $\omega_{c}$ is the cutoff
frequency), such as the multidimensional nonlinear Schr\"{o}dinger
(NLS) equation  and its generalizations, there is a completely
different type of vortex solitons (sometimes called spinning
solitons) having an intensity distribution in the form of a ring
or, in the 3D case, a torus, and these solitons can only be found
numerically. Such vortex solitons have been extensively studied in
BEC \cite{Saito2002,Carr2006,Parker2008}, nonlinear optics
\cite{Mihalache2005,Torner2005}, and, to a lesser extent, in a
plasma \cite{Berezhiani2010,Lashkin2020}. A distinctive feature of
such vortex solitons is their symmetry-breaking azimuthal
instability (snaking instability) (see the recent review
\cite{Malomed2019} and references therein).

A remarkable property of the Larichev-Reznik soliton is the
stability of solitons under head-on and overtaking collisions with
zero-impact parameter between the solitons
\cite{Makino1981,Williams1982}. In these cases the solitons
preserve their form after the collisions, and they behave just
like the one-dimensional solitons in the NLS equation and the KdV
equation  \cite{Ablowitz1981}. For the first time, the 3D
generalization of the Larichev-Reznik dipole solution was obtained
in \cite{Berestov1979,Berestov1981}. Recently, in the framework of
the 3D generalization of the Hasegawa-Mima equation, its the 3D
analytic soliton solutions were obtained and, as for the 2D
Larichev-Reznik solution, a remarkable elastic character of
collisions between the 3D solitons was demonstrated
\cite{Lashkin2017}.

The aim of this paper is to obtain a set of three-dimensional
nonlinear equations describing the dynamics of disturbances in a
self-gravitating rotating weakly inhomogeneous fluid system with
characteristic frequencies much lower than the rotation frequency,
that is, in the so-called geostrophic approximation. In a
particular long-wavelength case, when the characteristic size of
disturbances  is small compared to the Jeans length, we find the
3D analytical solutions of the corresponding equations in the form
of vortex dipole solitons.  Through numerical simulations, we show
that some of these 3D soliton solutions turn out to be extremely
stable.

The paper is organized as follows. In Sec. II, we present the
derivation of a set of nonlinear equations from the fluid
equations. In Sec. III, we consider the short-wavelength case and
present the 2D (pseudo 3D) soliton solutions. Sec. IV deals with
the long-wavelength case, where an analog of the 3D CHM equation
is obtained. In Sec. V, we obtain analytical solutions in the form
of three-dimensional vortex dipole solitons of various types. Sec.
VI is devoted to the study of stability of the found analytical
solutions. Finally, Sec. VII concludes the paper.

\section{ Model Equations}

Let us consider a gravitating system rotating with constant
angular velocity $\mathbf{\Omega}_{0}=\Omega_{0}\mathbf{\hat{z}}$
and with an equilibrium density $\rho_{0}$ in the plane
perpendicular to the $\mathbf{\hat{z}}$-axis. The momentum and
continuity fluid equations governing the dynamics of
self-gravitating rotating isothermal gas are
\begin{equation}
\label{momentum} \frac{\partial \mathbf{v}}{\partial
t}+(\mathbf{v}\cdot\nabla)\mathbf{v}=-\nabla\chi+2\Omega_{0}[\mathbf{v}\times\hat{\mathbf{z}}]
+\Omega_{0}^{2}[\hat{\mathbf{z}}\times[\mathbf{r}\times
\hat{\mathbf{z}}]],
\end{equation}
where the function $\chi$ is defined as
\cite{Fridman1991,Pokhotelov1998}
\begin{equation}
\label{chi}
\nabla\chi=\nabla\psi+\frac{c_{s}^{2}}{\rho}\nabla\rho,
\end{equation}
\begin{equation}
\label{continuity} \frac{\partial \rho}{\partial t}+\nabla\cdot
(\rho \mathbf{v})=0,
\end{equation}
where the first term in the right-hand side of Eq.
(\ref{momentum}) includes the pressure gradient force and
self-gravity force, second and third terms  account for the
Coriolis force and centrifugal force respectively. Here, $\rho$ is
the total mass density, $\mathbf{v}$ is the fluid velocity, $\psi$
is the gravitational potential, $c_{s}$ is the isothermal speed of
sound. Equations (\ref{momentum}) and (\ref{continuity}) are
supplemented by the Poisson equation for the gravity potential
$\psi$
\begin{equation}
\Delta\psi=4\pi G\rho \label{Poisson},
\end{equation}
where $G$ is the gravitational constant. We present the potential
and density as a sum of equilibrium and perturbed quantities
\begin{equation}
\label{perturbed} \psi=\psi_{0}+\tilde{\psi},\quad
\rho=\rho_{0}+\tilde{\rho}.
\end{equation}
At equilibrium we have
\begin{equation}
\label{equilib} \frac{\partial\chi_{0}}{\partial
r}=\Omega_{0}^{2}r ,
\end{equation}
whereas for the perturbations
\begin{equation}
\label{momentum1} \frac{\partial \mathbf{v}}{\partial
t}+(\mathbf{v}\cdot\nabla)\mathbf{v}=-\nabla\tilde{\chi}+2\Omega_{0}[\mathbf{v}\times\hat{\mathbf{z}}].
\end{equation}
From Eq. (\ref{Poisson}) we have
\begin{equation}
\label{Jeans} \Delta\psi_{0}=4\pi G\rho_{0}\equiv \omega_{0}^{2},
\quad \Delta\tilde{\psi}=4\pi G\tilde{\rho},
\end{equation}
where we have introduced the notation for the Jeans frequency
$\omega_{0}$. We assume a weak inhomogeneity of the equilibrium
density $\rho_{0}$ in the radial direction with a characteristic
inhomogeneity length $L$, so that all characteristic scales of
perturbations are much larger than $L^{-1}$, and use the local
Cartesian coordinate system ($x$ corresponds to the radial
coordinate $r$ and $y$ corresponds to the polar angle $\varphi$),
\begin{equation}
\label{equilib-density}
\omega_{0}^{2}(\mathbf{r})=\omega_{0}^{2}\left(1+\frac{x}{L}\right),
\end{equation}
where $x\ll L$. Substituting Eq. (\ref{perturbed}) into the
continuity equation (\ref{continuity}) and using Eq.
(\ref{Jeans}), one can obtain
\begin{equation}
\frac{\partial \Delta\tilde{\psi}}{\partial
t}+(\omega_{0}^{2}+\Delta\tilde{\psi})\nabla\cdot\mathbf{v}+\mathbf{v}\cdot\nabla
(\omega_{0}^{2}+\Delta\tilde{\psi})=0. \label{continuityPoisson}
\end{equation}
We assume that temporal variation of perturbations is slow
compared to the rotation frequency $\Omega_{0}$ and introduce the
ordering
\begin{equation}
\label{ordering} \epsilon\equiv \frac{\partial/\partial
t}{\Omega_{0}}\sim \frac{(\mathbf{v}\cdot\nabla)}{\Omega_{0}}\sim
\frac{\partial v_{z}/\partial z}{\Omega_{0}}\sim \frac{x}{L}.
\end{equation}
In the following we omit the tilde for the perturbed quantities.
Then, from the momentum equation  (\ref{momentum1}), taking into
account Eqs. (\ref{chi}) and (\ref{Jeans}), one can obtain to
lowest order in $\epsilon$ the velocity $\mathbf{v} _{\perp}$
perpendicular to the rotation axis
\begin{equation}
\label{v_0} \mathbf{v} _{\perp}^{(0)}
=\frac{1}{2\Omega_{0}}[\hat{\mathbf{z}}\times\nabla_{\perp}\Pi] ,
\end{equation}
where
\begin{equation}
\label{Pi} \Pi=\tilde{\psi}+\frac{c_{s}^{2}}
{\omega_{0}^{2}}\Delta\tilde{\psi}.
\end{equation}
To the next order we have
\begin{equation}
\label{v_1} \mathbf{v} _{\perp}^{(1)} =\mathbf{v}
_{\perp}^{(0)}-\frac{1}{4\Omega_{0}^{2}}\frac{d}{dt}\nabla_{\perp}\Pi
\end{equation}
where $d/dt=\partial/\partial t
+(\mathbf{v}_{\perp}^{(0)}\cdot\nabla)$. With this ordering, and
taking into account that $\nabla_{\perp}\cdot\mathbf{v}
_{\perp}^{(0)}=0$, we have from Eq. (\ref{continuityPoisson})
\begin{gather}
\frac{\partial \Delta\tilde{\psi}}{\partial
t}+\omega_{0}^{2}\left(\nabla_{\perp}\cdot\mathbf{v}
_{\perp}^{(1)}+\frac{\partial v_{z}}{\partial z}\right)+\mathbf{v}
_{\perp}^{(0)}\cdot\nabla_{\perp} \omega_{0}^{2}
 \nonumber \\
+\mathbf{v} _{\perp}^{(0)}\cdot\nabla_{\perp}\Delta\tilde{\psi}=0.
 \label{main11}
\end{gather}
The expression
$[\hat{\mathbf{z}}\times\nabla_{\perp}f]\cdot\nabla_{\perp}g=\{f,g\}$,
where $f$ and $g$ are arbitrary functions, also known as the
Poisson bracket defined by
\begin{equation}
\{f,g\}=\frac{\partial f}{\partial x}\frac{\partial g}{\partial y}
-\frac{\partial f}{\partial y}\frac{\partial g}{\partial x}.
\end{equation}
Then, substituting Eqs. (\ref{v_0}) and (\ref{v_1}) into Eq.
(\ref{main11}), one can obtain
\begin{equation}
 \label{main2}
\frac{\partial \Phi}{\partial
t}-\frac{\omega_{0}^{2}}{2\Omega_{0}L} \frac{\partial
\Pi}{\partial y}
+\frac{1}{2\Omega_{0}}\{\Pi,\Phi\}+\omega_{0}^{2}\frac{\partial
v_{z}}{\partial z}=0,
\end{equation}
where
\begin{equation}
 \label{Phi}
\Phi=\Delta\psi-\frac{\omega_{0}^{2}}{4\Omega_{0}^{2}}\Delta_{\perp}\Pi
.
\end{equation}
Here, and in what follows, the tilde is omitted for convenience.
Equation for the velocity along the rotation axis $v_{z}$, taking
into account Eqs. (\ref{v_0}) and (\ref{v_1}) with the ordering
Eq. (\ref{ordering}), follows from Eq. (\ref{momentum}) and has
the form
\begin{equation}
\label{main_z} \frac{\partial v_{z}}{\partial
t}+\frac{1}{2\Omega_{0}}\left\{\Pi,v_{z}\right\}+\frac{\partial
\Pi}{\partial z}=0.
\end{equation}
Equations (\ref{main2}) and (\ref{main_z}) are full set to
describe the dynamics of nonlinear perturbations. In the linear
approximation, taking $\psi (\mathbf{r},t)\sim \exp
(i\mathbf{k}\cdot\mathbf{r}-i\omega t)$ and
$v_{z}(\mathbf{r},t)\sim \exp (i\mathbf{k}\cdot\mathbf{r}-i\omega
t)$, Eqs. (\ref{main2}) and (\ref{main_z}) yield the dispersion
relation
\begin{equation}
\label{linear-general}
\omega^{2}\left[\frac{k^{2}}{(k^{2}-k_{J}^{2})}+\frac{k_{\perp}^{2}c_{s}^{2}}{4\Omega_{0}^{2}}\right]
+\omega\frac{k_{y}c_{s}^{2}}{2\Omega_{0}L} -k_{z}^{2}c_{s}^{2}=0,
\end{equation}
where $k^{2}=k^{2}_{\perp}+k^{2}_{z}$ with
$k_{\perp}^{2}=k_{x}^{2}+k_{y}^{2}$, $\omega$ is the frequency,
and $k_{J}=1/\lambda_{J}$, $\lambda_{J}=c_{s}/\omega_{0}$ is the
Jeans length. Then Eq. (\ref{linear-general}) predicts an
instability if
\begin{equation}
\label{stability_condition}
\frac{\omega_{0}^{2}}{4\Omega_{0}^{2}}\left(\frac{k_{y}^{2}}{L^{2}}+4k_{z}^{2}k_{\perp}^{2}\right)<
\frac{4k_{z}^{2}k^{2}}{1-k^{2}\lambda_{J}^{2}}.
\end{equation}
In particular, from Eq. (\ref{linear-general}) it follows that in
the stability region  there are two branches of oscillations: the
wave due to density inhomogeneity (if $\omega
k_{y}c_{s}^{2}/(2\Omega_{0}L)\gg k_{z}^{2}c_{s}^{2}$),
\begin{equation}
\label{branch1}
\omega=\frac{k_{y}c_{s}^{2}}{2\Omega_{0}L[k^{2}/(k^{2}-k^{2}_{J})
+k^{2}_{\perp}c_{s}^{2}/(4\Omega_{0}^{2})]},
\end{equation}
and the acoustic wave (if $\omega k_{y}c_{s}^{2}/(2\Omega_{0}L)\ll
k_{z}^{2}c_{s}^{2}$),
\begin{equation}
\label{branch2}
\omega=\frac{k_{z}c_{s}}{\sqrt{k^{2}/(k^{2}-k^{2}_{J})
+k^{2}_{\perp}c_{s}^{2}/(4\Omega_{0}^{2})}}.
\end{equation}
The classical Jeans instability condition in a homogeneous
non-rotating self-gravitating system, as is well known, has the
form $k\lambda_{J}<1$ , therefore, under the considered
conditions, the region of instability in terms of wave numbers
decreases significantly.

\section{Short-wavelength case and vortex tubes}

First we consider the short-wavelength perturbations with
$k\lambda_{J}\gg 1$. On the other hand, for sufficiently short
wavelengths, the hydrodynamic model is not valid and a kinetic
description is needed in the framework of the Vlasov kinetic
equation \cite{KineticJeans2004}. Therefore, we also require
$k\lambda_{D}\ll 1$, where $\lambda_{D}=v_{T}/\omega_{0}$ is the
Debye radius, $v_{T}$ being the thermal velocity of the small
gravitating masses \cite{KineticJeans2004}. We introduce
dimensionless variables as
\begin{equation}
\mathbf{r}\rightarrow
\frac{\lambda_{J}\omega_{0}}{2\Omega_{0}}\mathbf{r},\,
t\rightarrow \frac{t}{2\Omega_{0}}, \,
\tilde{\rho}=\frac{\rho_{0}\Delta\psi}{\omega_{0}^{2}}\rightarrow
\rho_{0}n, \, v_{z}\rightarrow c_{s}v_{z},
\end{equation}
where the variables on the left-hand side are physical variables
and those on the right-hand side are used subsequently. Then, from
Eqs. (\ref{Pi}) and (\ref{main2})--(\ref{main_z}), we have
\begin{equation}
\label{main4} \frac{\partial}{\partial
t}(n-\Delta_{\perp}n)-v_{\ast}\frac{\partial n}{\partial
y}-\{n,\Delta_{\perp}n\}+\frac{\partial v_{z}}{\partial z}=0,
\end{equation}
\begin{equation}
\label{main5} \frac{\partial v_{z}}{\partial
t}+\nu\{n,v_{z}\}+\frac{\partial n}{\partial z}=0,
\end{equation}
where $v_{\ast}=c_{s}/(2\Omega_{0}L)$. The system of equations
(\ref{main4}) and (\ref{main5}) is similar to the system of
equations obtained in Ref.~\cite{Horton1983} to describe nonlinear
drift waves in a plasma except for the sign in the term with
$v_{\ast}$. Following Ref.~\cite{Horton1983} and looking for
stationary traveling solutions to Eqs. (\ref{main4}) and
(\ref{main5}) of the form
\begin{gather}
n (x,y,z,t)=n (x,\xi), \\
v_{z} (x,y,z,t)=v_{z} (x,\xi),
\end{gather}
where $\xi=y-ut+\alpha z$ and  $u$ is the velocity of propagation
in the $y$ direction. Using Eq. (\ref{main5}), one can get
\begin{equation}
v_{z}(x,\xi)=\frac{\alpha }{u}n(x,\xi),
\end{equation}
and then from (\ref{main4}) it follows
\begin{equation}
\{n-ux,\Delta_{\perp}n-n+(\alpha^{2}/u-v_{\ast})x\}=0.
\end{equation}
Obtaining a localized solution of the resulting 2D nonlinear
equation using the Larichev-Reznik procedure is reduced to solving
two independent linear equations for the inner and outer (with a
circular cut in the plane) spatial regions, respectively. The
corresponding two solutions are matched at the cut boundary in
such a way that not only the solution itself of the original
nonlinear equation, but also all derivatives up to the second
order inclusive, must be continuous. The Larichev-Reznik method
(in addition to the original works
\cite{Larichev1976a,Larichev1976b} ) is described in detail in
many works
Refs.~\cite{Hasegawa1978,Flierl1980,Makino1981,Williams1982,Horton1983,Petviashvili_book1992}).
In the polar coordinates
\begin{equation}
x=r\cos\varphi, \,\, \xi=r\sin\varphi,
\end{equation}
the solution obtained in \cite{Horton1983} has the form
\begin{equation}
\label{Horton} n(r,\varphi)=ua\cos\varphi\left\{
\begin{array}{lc}
\displaystyle
\left(1+\frac{\beta^{2}}{\gamma^{2}}\right)\frac{r}{a}-\frac{\beta^{2}}{\gamma^{2}}\frac{J_{1}(\gamma
r/a)}{J_{1}(\gamma)}
,&  r\leqslant a , \\
\displaystyle  \,\frac{K_{1}(\beta r/a)}{K_{1}(\beta)},&
r\geqslant a ,
\end{array}
\right.
\end{equation}
where
\begin{equation}
\label{beta}
\beta=\sqrt{1+\frac{v_{\ast}}{u}-\frac{\alpha^{2}}{u^{2}}},
\end{equation}
and $\gamma$ is determined by
\begin{equation}
\frac{K_{2}(\beta)}{\beta
K_{1}(\beta)}=-\frac{J_{2}(\gamma)}{\gamma J_{1}(\gamma)}.
\end{equation}
In Eq. (\ref{Horton}), $J_{n}$ and $K_{n}$ are Bessel and McDonald
functions of order $n$. The solution is bounded at zero $r=0$ and
decreases exponentially at infinity, being an essentially
nonlinear solution in the form of a two-dimensional (pseudo
three-dimensional) soliton with embedded vorticity $(\nabla\times
\mathbf{v})_{z}\neq 0$ (a modon). It is a pair of vortices
rotating in the opposite direction, that is, a
cyclone-anticyclone. The modon solution (\ref{Horton}) has three
independent free parameters: the velocity $u$, the modon radius
$a$ (characteristic size), and $\alpha$ is the angle of
inclination of the vortex front with respect to the plane
perpendicular to the $z$-axis. As follows from Eq. (\ref{beta}),
the soliton velocity is limited by condition
\begin{equation}
\label{u-condition} u^{2}+v_{\ast} u-\alpha^{2}> 0.
\end{equation}
Within the interior region $r < a$, the fluid particles are
trapped and are thus transported along the direction of modon
movement. The density perturbation $n$ is continuous at the
boundary $r=a$ together with the first and second derivatives. For
$\alpha=0$, the solution (\ref{Horton}) reduces to the
Larichev-Reznik solution.

\section{Long-wavelength case}

Next, we consider the long-wavelength case, when the
characteristic lengths of perturbations are much less than the
Jeans length, i.e., $k\lambda_{J}\ll 1$. In this case, we obtain a
three-dimensional nonlinear equation, which, under certain
conditions, admits analytical 3D soliton solutions. In this case,
from Eqs. (\ref{Pi}) and (\ref{main2})--(\ref{main_z})
 we have
\begin{gather}
\frac{\partial}{\partial
t}\left(\Delta\psi-\nu\Delta_{\perp}\psi\right)
-\frac{2\Omega_{0}\nu}{L}\frac{\partial\psi}{\partial
y}+\frac{1}{2\Omega_{0}}\left\{\psi,\Delta\psi-\nu\Delta_{\perp}\psi\right\}
\nonumber \\
 +\omega_{0}^{2}\frac{\partial v_{z}}{\partial z}=0,
 \label{basic5}
\end{gather}
and
\begin{equation}
 \label{basic6}
\frac{\partial v_{z}}{\partial
t}+\frac{1}{2\Omega_{0}}\left\{\psi,v_{z}\right\}+\frac{\partial\psi}{\partial
z}=0,
\end{equation}
where $\nu=\omega_{0}^{2}/(4\Omega_{0}^{2})$. In the linear
approximation, the dispersion relation is
\begin{equation}
\omega^{2}(k^{2}-\nu
k_{\perp}^{2})-2\omega\frac{k_{y}\Omega_{0}\nu}{L}+k_{z}^{2}\omega_{0}^{2}=0.
\end{equation}
Neglecting parallel motion, i.e. the interaction with the acoustic
branch of oscillations, from Eqs. (\ref{basic5}) and
(\ref{basic6}) one can obtain
\begin{equation}
\label{without-parallel} \frac{\partial}{\partial
t}(\Delta\psi-\nu\Delta_{\perp}\psi)-\frac{2\Omega_{0}\nu}{L}\frac{\partial\psi}{\partial
y}+\frac{1}{2\Omega_{0}}\left\{\psi,\Delta\psi-\nu\Delta_{\perp}\psi\right\}=0.
\end{equation}
The existence of stationary solutions of Eq.
(\ref{without-parallel}) requires that the operator
$\Delta-\nu\Delta_{\perp}$, depending on the spatial derivatives,
be elliptic. One can easily see that this leads to the condition
$\nu < 1$. In this paper, we restrict ourselves to just this case.
Next, we introduce dimensionless variables
$\mathbf{r}_{\perp}^{\prime}$, $z^{\prime}$,  $t^{\prime}$, and
$\psi^{\prime}$ by
\begin{equation}
\label{dimensionless} \mathbf{r}_{\perp}=
\frac{L\sqrt{1-\nu}}{2\nu}\mathbf{r}_{\perp}^{\prime},\,\, z=
\frac{L}{2\nu}z^{\prime},\,\, t\rightarrow
\frac{t^{\prime}}{\Omega_{0}}, \,\, \psi=2\Omega_{0}^{2}
\psi^{\prime},
\end{equation}
and further the primes are omitted. Substituting Eq.
(\ref{dimensionless}) into Eq. (\ref{without-parallel}) we have
\begin{equation}
\label{HM3D} \frac{\partial \Delta\psi}{\partial
t}-w\frac{\partial\psi}{\partial
y}+\left\{\psi,\Delta\psi\right\}=0,
\end{equation}
where $w=1/\sqrt{1-\nu}$, and Eq. (\ref{HM3D}) can be rewritten as
\begin{equation}
\label{HM3D1} \frac{\partial \Gamma}{\partial
t}+\left\{\psi,\Gamma\right\}=0,
\end{equation}
where $\mathbf{v}_{D}=[\hat{\mathbf{z}}\times\nabla_{\perp}\psi]$,
and $\Gamma=\Delta\psi +wx$ is the generalized vorticity, or,
equivalently, as
\begin{equation}
\label{G_t1} \frac{\partial \Gamma}{\partial
t}+\mathbf{v}_{D}\cdot\nabla\Gamma=0.
\end{equation}
Note that a three-dimensional generalization of the Charney
equation in geophysics was first obtained in
Refs.~\cite{Berestov1979,Berestov1981}, and the Hasegawa-Mima
equation for plasma in Ref.~\cite{Lashkin2017}. Equation
(\ref{HM3D}) differs from the 3D Charney-Hasegawa- Mima equation
obtained earlier in \cite{Berestov1979,Berestov1981,Lashkin2017}
by the absence of an additional term $-\partial\psi/\partial t$.
Equation (\ref{G_t1}) describes the generalized vorticity
convection in an incompressible velocity field $\mathbf{v}_{D}$
with $d\Gamma/dt=0$, where $d/dt=\partial/\partial
t+\mathbf{v}_{D}\cdot\nabla_{\perp}$. Like Eq. (19) in
Ref.~\cite{Lashkin2017}, Eq. (\ref{HM3D}) and has an infinite set
of integrals of motion (Casimir invariants),
\begin{equation}
\label{integral1} \int f(\Gamma,z)\,d^{3}\mathbf{r},
\end{equation}
where $f$ is an arbitrary function of its arguments. Other
integrals of motion are
\begin{equation}
\label{integral2} \int \psi\Gamma\,d^{3}\mathbf{r}, \quad \int
x\Gamma\,d^{3}\mathbf{r}, \quad  \int
(y+v_{\ast}t)\Gamma\,d^{3}\mathbf{r}.
\end{equation}
The energy $E$ and enstrophy $K$, which are quadratic invariants,
coincide with the energy and enstrophy of the equation obtained in
Ref.~\cite{Lashkin2017},
\begin{gather}
\label{integrals_E_K}  E=\int
\left[\psi^{2}+(\nabla\psi)^{2}\right]\,d^{3}\mathbf{r}   , \\
K=\int
\left[(\nabla\psi)^{2}+(\Delta\psi)^{2}\right]\,d^{3}\mathbf{r}.
\end{gather}
As was pointed out in Ref.~\cite{Lashkin2017}, the presence of
such an infinite set of integrals of motion does not mean the
complete integrability of Eq. (\ref{HM3D}), just like the
two-dimensional CHM equation \cite{Shulman1988}.

\section{Three-dimensional vortex solitons}

We look for stationary traveling wave solutions of Eq.
(\ref{HM3D}) of the form
\begin{equation}
\label{mov_reference} \psi(x,y,z,t)=\psi(x,y',z), \, \, \,
y'=y-ut,
\end{equation}
where $u$ is the velocity of propagation in the $y$ direction (in
the following we omit the prime). Substituting Eq.
(\ref{mov_reference}) into Eq. (\ref{HM3D}), we have the relation
\begin{equation}
\label{stat_forma} \left\{\Gamma,\psi-ux\right\}=0,
\end{equation}
from which we can conclude that
\begin{equation}
\label{gen_solution} \Gamma=F(\psi-ux,z),
\end{equation}
where $F$ is an arbitrary function of both arguments. Following
the known procedure for finding modon solutions
\cite{Larichev1976a,Larichev1976a,Hasegawa1978,Flierl1980,Makino1981,Williams1982},
we assume that the generalized vorticity $\Gamma$ and stream
function $\psi$ satisfy one linear relation inside a region of
trapped fluid, and a different one outside, that is, $F$ is
piecewise linear function. For the 3D modon solutions
\cite{Berestov1979,Berestov1981,Flierl1987,Lashkin2017} the
trapped region is a sphere of radius $a$, and we have
\begin{equation}
\label{new-brace} F=\Delta\psi+wx=
\begin{cases}
 \displaystyle c_{1}(\psi-ux)+c_{2}+c_{3}z ,&  r < a
, \\  \displaystyle \,c_{4}(\psi-ux)+c_{5}+c_{6}z,& r
> a .
\end{cases}
\end{equation}
Note, that the linearity of function $F$ in the exterior region
$r>a$ follows from the requirement that the solution be localized
at infinity. Then it is easy to see that $c_{1}=-w/u$, $c_{2}=0$,
$c_{3}=0$ and we should have $u<0$, that is the solution moves in
the negative direction of the $y$-axis. The boundedness
requirement at $r=0$ implies $c_{4}<0$. In the following we
introduce the notations
\begin{equation}
\label{beta_gamma} \varkappa=a\sqrt{-w/u}, \, k=a\sqrt{-c_{4}}.
\end{equation}
Then Eq. (\ref{new-brace}) in the exterior and inner regions
become
\begin{equation}
\label{equ-ext}
 \Delta\psi-\frac{\varkappa^{2}}{a^{2}}\psi=0,
\end{equation}
and
\begin{equation}
\label{equ-in}
 \Delta\psi+\frac{k^{2}}{a^{2}}\psi=\frac{(\varkappa^{2}+k^{2})ux}{a^{2}}+c_{5}+c_{6}z,
\end{equation}
respectively. Equation (\ref{equ-ext}) has a general solution
\begin{equation}
\psi=\sum_{n,l,m}A_{nlm}\frac{K_{n+1/2}(\varkappa r/a)}{\sqrt{
r}}Y_{l}^{m}(\theta,\varphi),
\end{equation}
while a general solution of Eq. (\ref{equ-in}) with zero right
hand side is
\begin{equation}
\psi=\sum_{n,l,m}B_{nlm}\frac{J_{n+1/2}(k r/a)}{\sqrt{
r}}Y_{l}^{m}(\theta,\varphi),
\end{equation}
where we use spherical coordinates $(r,\theta,\varphi)$,
\begin{equation}
x=r\sin\theta\cos\varphi,\, y=r\sin\theta\sin\varphi,\,
z=r\cos\theta,
\end{equation}
and $n,m,l$ are integers, $J_{\nu}(\xi)$ is the Bessel function of
the first kind, $K_{\nu}(\xi)$ is the modified Bessel function of
the second kind, $Y_{lm}$  are the spherical harmonics, $A_{nlm}$
and $B_{nlm}$ are arbitrary constants. At present we consider only
the lowest radial modes $n=0,1$, and the lowest spherical
harmonics consistent with the terms $wx$ and $c_{6}z$, $l=0,1$,
and $m=0,\pm 1$. Then the real solution of Eq. (\ref{equ-ext}) for
the exterior region $r > a$ can be written as
\begin{gather}
\label{ps1} \psi=A_{000}\frac{K_{1/2}(\varkappa r/a)}{\sqrt{
r}}Y_{0}^{0}+A_{110}\frac{K_{3/2}(\varkappa r/a)}{\sqrt{
r}}Y_{1}^{0} \\ \nonumber +A_{111}\frac{K_{3/2}(\varkappa
r/a)}{\sqrt{ r}}(Y_{1}^{1}+Y_{1}^{-1}),
\end{gather}
where the unnormalized spherical harmonics have the form
$Y_{0}^{0}=1$, $Y_{1}^{0}=\cos\theta$ and $Y_{1}^{\pm 1}=\exp (\pm
i\varphi)\sin\theta$. A general solution of Eq. (\ref{equ-in}) for
the inner region is the sum of the general solution of the
corresponding homogeneous equation and the particular solution of
the complete inhomogeneous equation. As a particular solution, it
is easy to see that we can take
\begin{equation}
\psi_{par}=\left(1+\frac{\varkappa^{2}}{k^{2}}\right)ur\sin\theta\cos\varphi+\frac{c_{5}a^{2}}{k^{2}}
+\frac{c_{6}a^{2}}{k^{2}}\cos\theta .
\end{equation}
Then, a general solution of Eq. (\ref{equ-in}) for the inner
region $r < a$ has the form
\begin{gather}
\label{ps2} \psi=B_{000}\frac{J_{1/2}(\varkappa r/a)}{\sqrt{
r}}Y_{0}^{0}+B_{110}\frac{J_{3/2}(\varkappa r/a)}{\sqrt{
r}}Y_{1}^{0} \\ \nonumber +B_{111}\frac{J_{3/2}(\varkappa
r/a)}{\sqrt{ r}}(Y_{1}^{1}+Y_{1}^{-1})+\psi_{par}.
\end{gather}
We require that $\psi$ and $\nabla\psi$ to be continues at $r=a$
\begin{equation}
\label{cond1} \psi\mid_{r=a-0}=\psi\mid_{r=a+0}, \, \,
\nabla\psi\mid_{r=a-0}=\nabla\psi\mid_{r=a+0},
\end{equation}
and $\Delta\psi$ (or, equivalently, $\Gamma$) has a constant jump
$p$ (including the case $p=0$) at $r=a$
\begin{equation}
\label{cond2} \Delta\psi\mid_{r=a-0}=\Delta\psi\mid_{r=a+0}+p.
\end{equation}
The presence of such a jump leads, just as for the modon in
Ref.~\cite{Lashkin2017}, to the appearance of a radially symmetric
part in the solution. By substituting Eqs. (\ref{ps1}) and
(\ref{ps2}) into Eqs. (\ref{cond1}) and (\ref{cond2}), we can find
the desired solution. In this case, for a given value of
$\varkappa$, the value of $k$ is determined by the relation
\begin{equation}
\label{transzent} (k^{2}\delta+3-k^{2})\tan k=k(k^{2}\delta+3) \,
,
\end{equation}
where
\begin{equation}
\label{delta}
\delta=\frac{(\varkappa^{2}+3\varkappa+3)}{\varkappa^{2}(\varkappa+1)}.
\end{equation}
Given that the functions $J_{n+1/2}$ and $K_{n+1/2}$ for integer
values of the index $n$ can be expressed in terms of trigonometric
functions and the exponential function, respectively (together
with rational ones), the final solution can be written as
\begin{equation}
\label{final_solution}
\psi(r,\theta,\varphi)=\Psi_{0}(r)+\Psi(r)(\sin\theta\cos\varphi+\mu\cos\theta),
\end{equation}
where $\mu$ is an arbitrary constant, and $\Psi_{0}(r)$ and
$\Psi(r)$ are determined by
\begin{widetext}
\begin{equation}
\label{psi0}
\Psi_{0}(r)=\frac{pa^{2}}{(\varkappa^{2}+k^{2})\delta}\left\{
\begin{array}{lc}
\displaystyle  \, \frac{a\sin (k r/a)}{r(\sin k-k\cos
k)}-\frac{3(\varkappa^{2}+k^{2})}{\varkappa^{2}k^{2}},
&  r\leqslant a \\
\displaystyle
\,\frac{a}{(1+\varkappa)r}\exp\left[-\varkappa\left(\frac{r}{a}-1\right)\right],&
r\geqslant a
\end{array}
\right. ,
\end{equation}
\end{widetext}
and
\begin{widetext}
\begin{equation}
\label{psi} \Psi(r)=ua\left\{
\begin{array}{lc}
\displaystyle
\left(1+\frac{\varkappa^{2}}{k^{2}}\right)\frac{r}{a}-
\frac{\varkappa^{2}}{k^{2}}\frac{a^{2}[\sin (k
r/a)-(kr/a)\cos(kr/a)]}{r^{2}(\sin k-k\cos k)}
,&  r\leqslant a \\
\displaystyle  \,\frac{a^{2}(1+\varkappa
r/a)}{r^{2}(1+\varkappa)}\exp\left[-\varkappa\left(\frac{r}{a}-1\right)\right],&
r\geqslant a
\end{array}
\right. .
\end{equation}
\end{widetext}
As can be seen from Eqs. (\ref{final_solution}), (\ref{psi0}) and
(\ref{psi}), the 3D soliton consists of an $x$-antisymmetric
dipole part, on which, as on a carrier, there is a core - a
radially symmetric part of an arbitrary amplitude and a
$z$-antisymmetric dipole part of an arbitrary amplitude. The
carrier amplitude is determined by the velocity $u$ and
localization size of the soliton $a$. The core and the
$z$-antisymmetric parts cannot exist without the carrier. The
radially symmetric part vanishes if $p=0$, and, under this,
$\Delta\psi$ (and the vorticity) is continuous at the boundary
$r=a$. The $z$-antisymmetric part vanishes if $\mu=0$. Thus, the
3D soliton solution (\ref{final_solution}) has four independent
free parameters - the velocity $u$, the soliton characteristic
size
 $a$, value characterizing the amplitude of the $z$-antisymmetric part $\mu$, and
the jump of the vorticity $p$ which determines the amplitude of
the radially symmetric part. Within the interior region $r<a$, the
fluid particles are trapped and are thus transported along the
$y$-direction. In the exterior region $r>a$, the solution decays
exponentially to zero. Note that Eq. (\ref{transzent}) has an
infinite set of roots $k_{n}$, $n=1,2\dots$ for each $\varkappa$.
Therefore, Eqs. (\ref{final_solution}), (\ref{psi0}) and
(\ref{psi}) present the infinite set of solutions with $k=k_{n}$.
The solution with $n=1$ (the ground state) has no radial nodes.
The higher states have $n-1$ nodes (in the interior region). In
what follows, we consider only the ground state $n=1$. Note that,
as follows from Eq. (\ref{dimensionless}), the values $\nu\lesssim
1$ in physical variables correspond to the oblateness of the
soliton along the axis of rotation. In the limiting case
$\varkappa\rightarrow 0$, one can obtain
%\begin{widetext}
\begin{equation}
\label{psi0_bet0} \Psi_{0}(r)=\frac{pa^{2}}{k^{2}}\left\{
\begin{array}{lc}
\displaystyle \, \frac{a\sin (kr/a)}{r\sin k}-1,
&  r\leqslant a \\
\displaystyle \,0,& r\geqslant a
\end{array}
\right. ,
\end{equation}
and
\begin{widetext}
\begin{equation}
\label{psi_bet0} \Psi(r)=wa\left\{
\begin{array}{lc}
\displaystyle \frac{r}{a}- \frac{3a^{2}[\sin (k r/a)-(k r/a)\cos(k
r/a)]}{r^{2}k^{2}\sin k}
,&  r\leqslant a \\
\displaystyle  \,\frac{a^{2}}{r^{2}},& r\geqslant a
\end{array}
\right. ,
\end{equation}
\end{widetext}
Note that in this limiting case, the solution has a long tail,
that is, it decreases at infinity in a power-law manner, rather
than exponentially (which can be seen immediately from Eq.
(\ref{equ-ext})). In the other limiting case,
$\varkappa\rightarrow \infty$, that is $u\rightarrow 0$, which
corresponds to a motionless soliton, the radially symmetric
component disappears completely, $\Psi_{0}(r)=0$, and
\begin{widetext}
\begin{equation}
\label{psi_betinft} \Psi(r)=\frac{a^{3}w}{k^{2}}\left\{
\begin{array}{lc}
\displaystyle \frac{a^{2}[\sin (k r/a)-(k r/a)\cos(k
r/a)]}{r^{2}(\sin k-k\cos k)}-\frac{r}{a}
,&  r\leqslant a \\
\displaystyle  \,0,& r\geqslant a
\end{array}
\right. .
\end{equation}
\end{widetext}
so that the solution is completely screened in the outer region
$r>a$, although it remains continuous, as can be seen, up to the
second derivatives.

\begin{figure}
\includegraphics[width=3.4in]{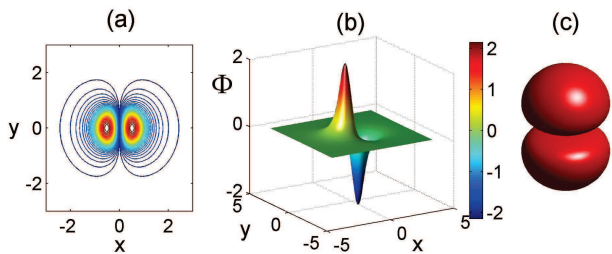}
\caption{\label{fig1} The 3D vortex soliton (\ref{final_solution})
with the parameters $u=-1$ (velocity), $a=1$ (cut radius), $p=0$
(no monopole part) and $\mu=0$ (no $z$-antisymmetric part): (a)
Streamlines $|\psi|$ in the $x-y$-plane section; (b) the field
$\psi$ in the $x-y$-plane section; (c) isosurface $|\psi
(x,y,z)|=0.8$. }
\end{figure}

\begin{figure}
\includegraphics[width=3.4in]{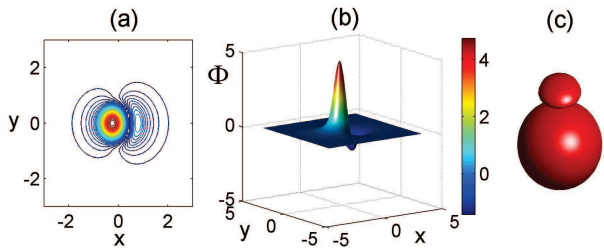}
\caption{\label{fig2} The 3D vortex soliton (\ref{final_solution})
with the parameters $u=-1.$ (velocity), $a=1$ (cut radius),
$p=-15$ (monopole part) and $\mu=0$ (no $z$-antisymmetric part):
(a) Streamlines $|\psi|$ in the $x-y$ plane section; (b) the field
$\psi$ in the $x-y$ plane section; (c) isosurface $|\psi
(x,y,z)|=1.3$. One can see how the monopole part masks the dipole
part.}
\end{figure}

\begin{figure}
\includegraphics[width=3.2in]{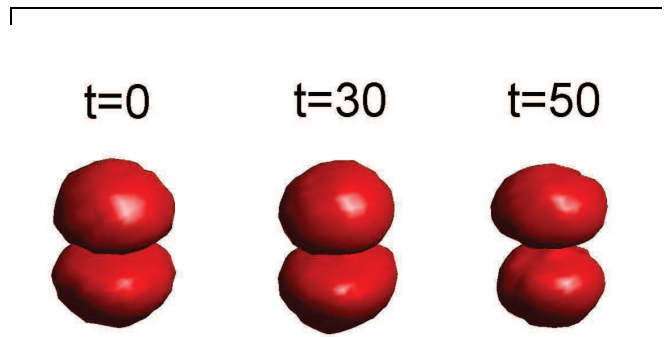}
\caption{\label{fig3} Evolution of the 3D soliton with the
parameters $u=-0.2$ (velocity), $a=1$ (cut radius), $p=0$ (no
monopole part) and $\mu=0$ (no $z$-antisymmetric part) in the
presence of strong initial random perturbation; the shape
(isosurface $|\psi (x,y,z)|=0.15$) of the perturbed soliton at the
initial moment,$t=0$, at $t=30$ and at $t=50$.}
\end{figure}

\begin{figure}
\includegraphics[width=3.2in]{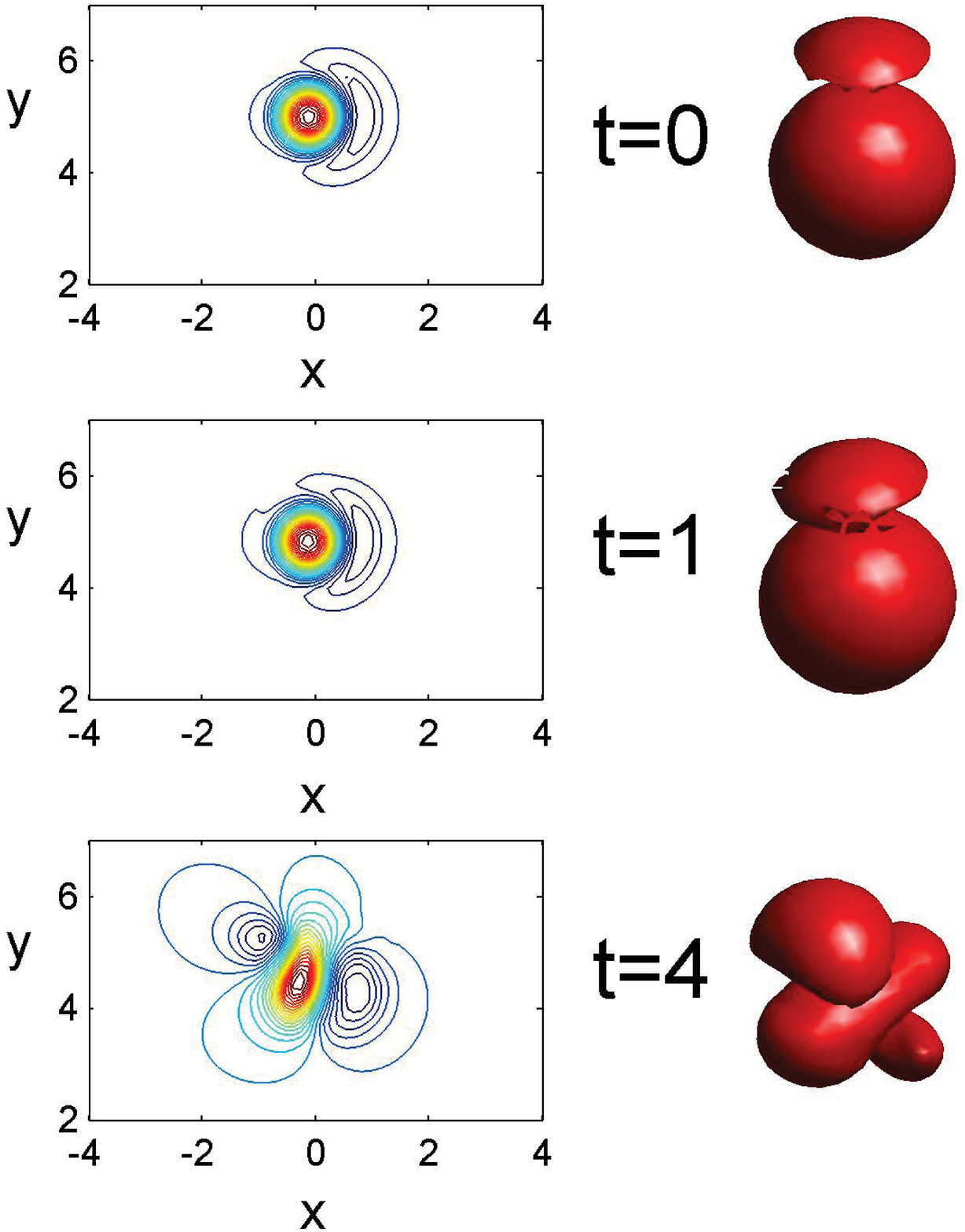}
\caption{\label{fig4} Fast destruction of the 3D soliton with a
large amplitude of the monopole part and the parameters $u=-0.2$
(velocity), $a=1$ (cut radius), and $p=-10$; isosurface $|\psi
(x,y,z)|=0.35$ is shown. }
\end{figure}

\begin{figure}
\includegraphics[width=3.2in]{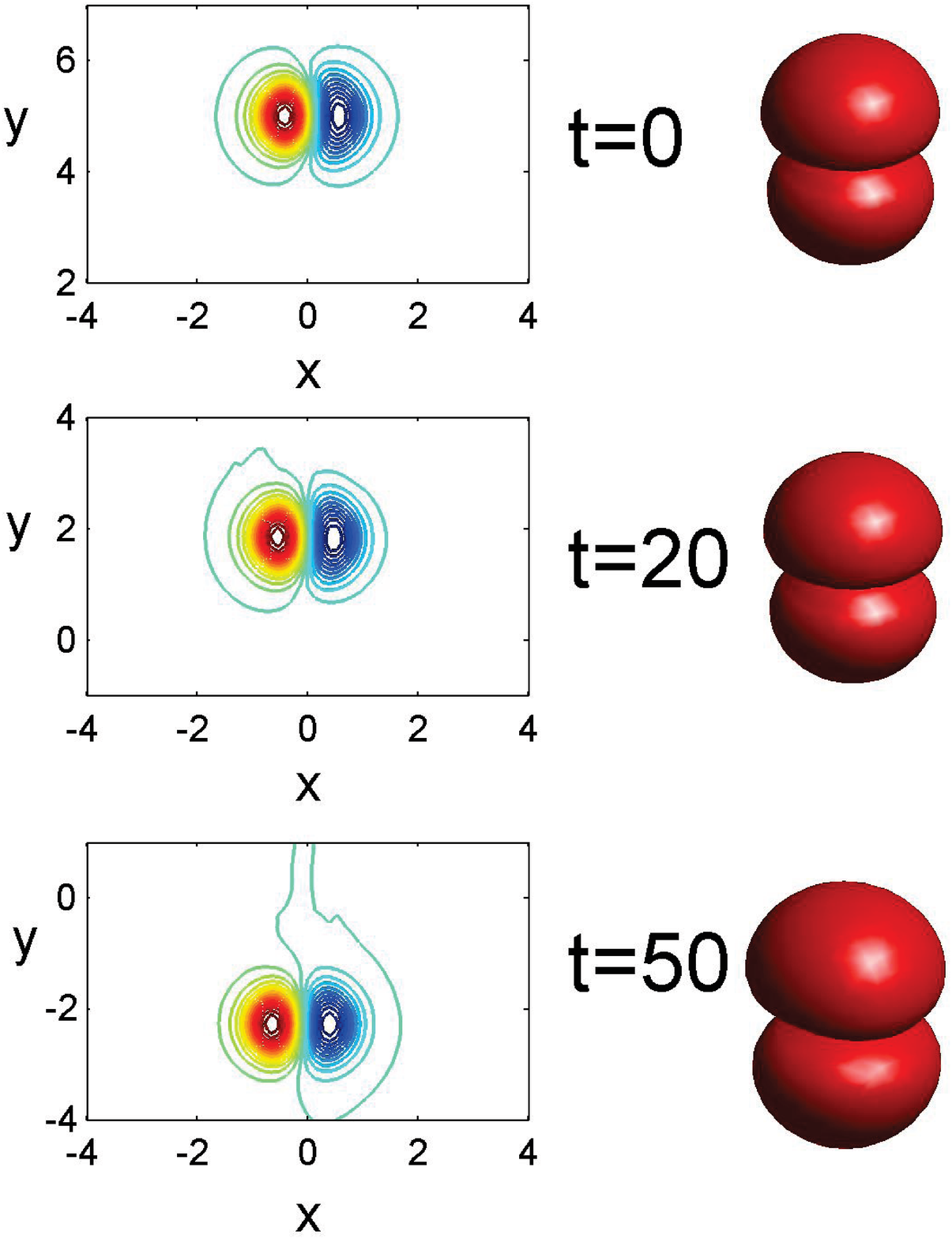}
\caption{\label{fig5} Almost stable 3D soliton dynamics with a
sufficiently small monopole part and the parameters $u=-0.2$
(velocity), $a=1$ (cut radius), and $p=-1$; isosurface $|\psi
(x,y,z)|=0.15$ is shown. }
\end{figure}

\begin{figure}
\includegraphics[width=3.2in]{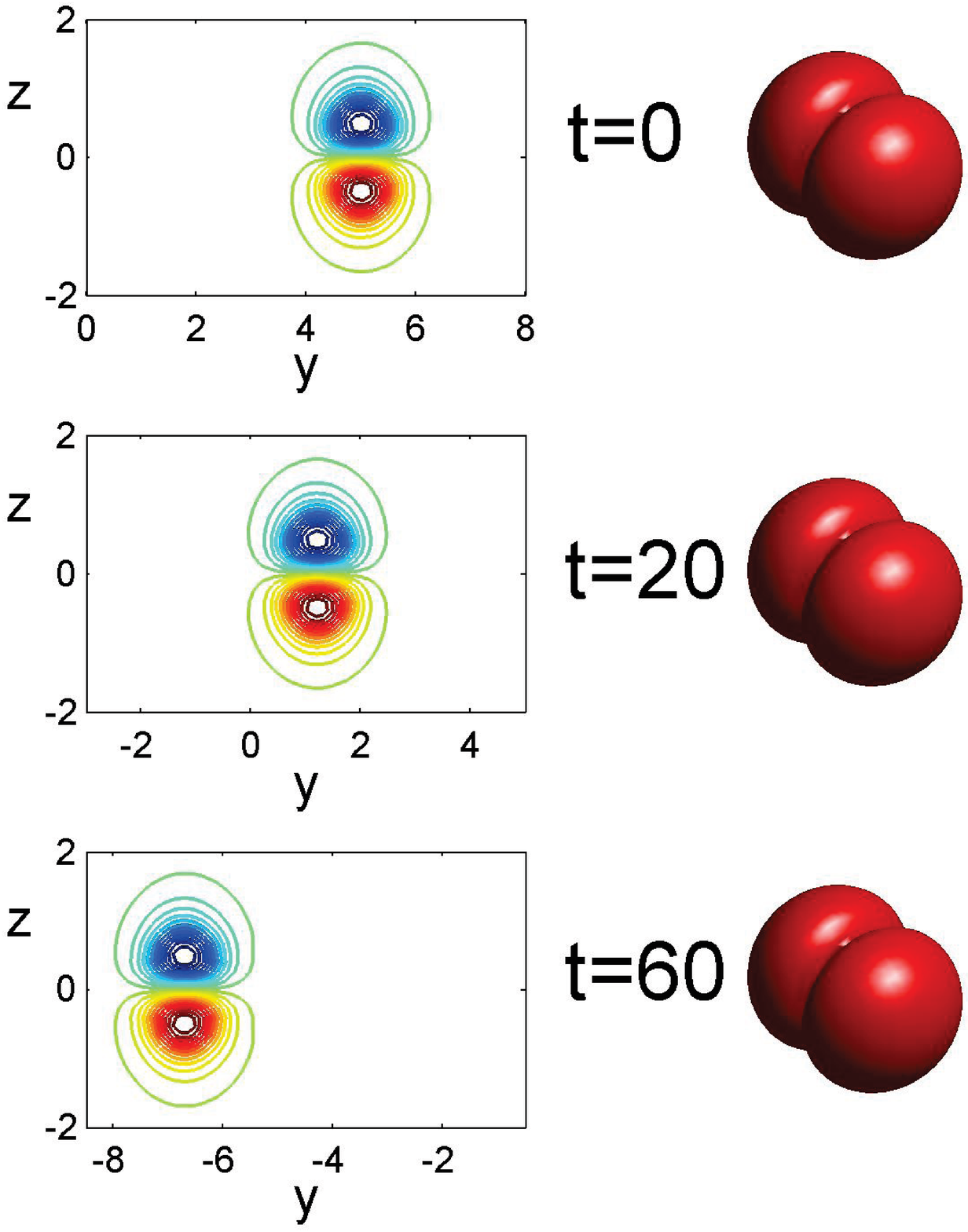}
\caption{\label{fig6} Stable 3D soliton dynamics with the
parameters $u=-0.2$ (velocity), $a=1$ (cut radius), $p=0$ and
$\mu=10$; isosurface $|\psi (x,y,z)|=0.3$ is shown. }
\end{figure}
We emphasize that, as noted above, the amplitude of the radially
symmetric part, determined by parameter $p$, can be arbitrary and
significantly exceed the amplitude of the basic part (carrier). In
this case, the monopole part masks the basic part  and the soliton
looks like a monopole vortex soliton. The 3D vortex soliton
solution (\ref{final_solution}) without a radially symmetric part
($p=0$) and without an $z$-antisymmetric part ($\mu=0$), moving
with the velocity $u=-1$ and  cut radius $a=1$, is shown in
Fig.~\ref{fig1}. From the given values $u$ and $a$, the value $k$
in Eqs. (\ref{final_solution}) and (\ref{psi}) is determined by
the numerical solution of the transcendental equation
(\ref{transzent}) and the smallest value $k$ corresponding to the
ground state is taken. For definiteness, we take $\nu=0.2$ here
and in all subsequent numerical simulations. Other values  $\nu$
do not qualitatively change the results. The soliton with the same
parameters, but with a radially symmetric part of a sufficiently
large amplitude with $p=-15$, is presented in Fig.~\ref{fig2}. In
this case, one can see a distinct monopole part.

\section{Stability of the 3D solitons}

To study the stability of the found exact 3D solutions, we
numerically solved the dynamical equation (\ref{HM3D}) with the
initial conditions corresponding to analytical solution
(\ref{final_solution}). The time integration is performed by an
implicit Adams-Moulton method with the variable time step and the
variable order and local error control (we used the corresponding
NAG (Numerical Algorithms Group) routine \cite{NAG18}). Periodic
boundary conditions are assumed. The linear terms are computed in
spectral space. The Poisson bracket nonlinearity is evaluated in
physical space by a finite difference method, using the energy-
and enstrophy-conserving Arakawa scheme \cite{Arakawa1966}
modified for the 3D case (see Appendix).

As a first (and principal) example, we consider the stability of
the 3D vortex soliton without superimposed parts, that is, without
an additional radially symmetric part, $\Psi_{0}=0$, and without
an additional antisymmetric part in $z$-axis, $\mu=0$, in Eq.
(\ref{final_solution}). The initial condition at the time $t=0$
was taken in the form
$\psi(\mathbf{r},0)=\psi_{s}(\mathbf{r},0)[1+\epsilon\xi
(\mathbf{r})]$, where $\psi_{s}=\Psi (r)\sin\theta\cos\varphi$,
and $\Psi (r)$ is determined by Eq. (\ref{psi}), $\xi
(\mathbf{r})$ is the white Gaussian noise with variance
$\sigma^{2}=1$, and the parameter of perturbation
$\epsilon=0.01-0.1$. The stable dynamics of such a vortex soliton
with $\epsilon=0.05$ is shown in Fig.~\ref{fig3}. It can be seen
that at the initial time $t=0$ the soliton is perturbed by a
sufficiently strong noise, however, the soliton, generally
speaking, does not undergo any significant shape distortions at
times $t=30$ and $t=50$. In particular, there is no fast
development of symmetry breaking between the cyclone and
anticyclonic parts. Note that there are two characteristic times
of the processes in the model: the dispersive time $\sim
1/\omega_{\mathbf{k}}$, where $\omega_{\mathbf{k}}$ is the linear
dispersion law, at which the packet of linear waves spreads out
due to dispersion, and the nonlinear time $\sim 1/\omega_{NL}$,
where $\omega_{NL}$ is the characteristic nonlinear frequency
(vortex rotation frequency) defined as
$\omega_{NL}=\mathbf{k}\cdot\mathbf{v}$, where $k\sim 1/a$ and
$\mathbf{v}_{\mathbf{k}}\sim
[\hat{\mathbf{z}}\times\mathbf{k}]\psi_{\mathbf{k}}$ is the fluid
velocity in the vortex. It can be seen from Fig.~\ref{fig3} that
for the given parameters of the vortex soliton, it evolves without
significant distortion of its shape over many periods of rotation
$T=2\pi /\omega_{NL}$. In the absence of initial noise
$\epsilon=0$, the soliton evolves for an arbitrarily long time
without any shape distortion (simulation was carried out up to
times $t=400$, but in fact we did not observe any distortions at
these times, for example, at $\epsilon=0.01$). Such behavior was
observed for various values of the soliton velocity $u$ and the
parameter $a$.

Then we studied the stability of solution (\ref{final_solution})
with a radially symmetric (monopole) part at different values of
the amplitude of this part. Recall that the magnitude of this
amplitude determines the magnitude of the vorticity jump $p$ at
the cut boundary $a$. We also assume that there is no additional
$z$-antisymmetric part. Here we assume that there is no additional
initial noise disturbance, and then the initial condition for Eq.
(\ref{HM3D}) is
$\psi(\mathbf{r},0)=\Psi_{0}(r)+\Psi(r)\sin\theta\cos\varphi$,
where $\Psi_{0}(r)$ and $\Psi(r)$ are determined by Eqs.
(\ref{psi0}) and (\ref{psi}), respectively. Numerical simulation
shows that this solution is unstable. With a sufficiently large
amplitude of the radially symmetric part (and, accordingly, the
vorticity jump $p$), such a soliton is destroyed almost
immediately. The destruction of the soliton with $p=-10$ is shown
in Fig.~\ref{fig4}. On the other hand, as seen in Fig.~\ref{fig5},
if the amplitude of the monopole part is not too large and $p=-1$,
the soliton retains its original shape for quite a long time,
although the appearance of an insignificant radiated wave wake is
already noticeable at the time $t=50$.

Finally, in  Fig.~\ref{fig6} shows the dynamics of a soliton
without a radially symmetric part, but with an additional
$z$-antisymmetric part of a sufficiently large amplitude with
$\mu=10$. Once again we emphasize that for such solutions all
second derivatives and vorticity are continuous. It can be seen
that the soliton moves without any distortion of its shape at
times $t=20$ and $t=60$. Nevertheless, such solitons, in contrast
to the solitons with $p=0$ and $\mu=0$, turn out to be unstable.
For parameter values $\mu=40$, the soliton is destroyed
(corresponding figure is not shown).

Note that the behavior of all three types of solitons considered
is consistent with the results of Ref.~\cite{Lashkin2017}, where
head-on and overtaking collisions between three-dimensional vortex
solitons were studied in a similar model. As noted above, in
contrast to Ref.~\cite{Lashkin2017}, the 3D vortex solitons of Eq.
(\ref{HM3D}) can move only in one direction.

\section{Conclusion}

In this paper, we have derived the system of 3D nonlinear
equations describing the dynamics of disturbances in a weakly
nonuniform rotating self-gravitating fluid under the assumption
that the characteristic frequencies of disturbances are small
compared to the rotation frequency. The nonlinear terms in this
system have the form of the Poisson bracket, which is quite common
in problems of nonlinear geophysics. Linear dispersion is due to
weak inhomogeneity. In a linear approximation, we obtained an
instability criterion and showed that the region of instability in
terms of wave numbers expands significantly in comparison with the
classical Jeans criterion for a homogeneous nonrotating system. In
the case when the characteristic perturbation lengths are much
larger than the Jeans length, the resulting system of nonlinear
equations can be reduced to the system previously obtained in
\cite{Horton1983}. The analytical solution of this system is
obtained by the well-known Larichev-Reznik method for finding the
2D nonlinear dipole solutions in the physics of atmospheres of
rotating planets and is actually the 2D Larichev-Reznik vortex
dipole soliton in the form of a cyclone-anticyclone pair (in fact,
the solution is a pseudo three-dimensional vortex tube). In the
opposite long-wavelength case of small perturbation lengths
compared to the Jeans length, we obtained the original 3D
nonlinear equation resembling the 3D analog of the 2D CHM
equation. We have obtained analytical the 3D soliton solutions of
this equation by generalizing the Larichev-Reznik procedure to the
3D case. The solution is a vortex soliton moving with the constant
velocity in the direction perpendicular to the direction of the
inhomogeneity and the direction of the axis of rotation. The main
part of the solution is $x$-antisymmetric, that is, along the
direction of the inhomogeneity, and is a three-dimensional dipole
in the form of a cyclone-anticyclone pair. In addition to the
basic 3D $x$-antisymmetric part, the solution may also contain
radially symmetric (monopole) or/and antisymmetric along the
rotation axis ($z$-axis) parts with arbitrary amplitudes. It is
important to note that these superimposed parts cannot exist
without the main part (carrier), although the amplitudes of these
superimposed parts can significantly exceed the amplitude of the
carrier. For example, if the amplitude of the radially symmetric
part is much greater than the amplitude of the antisymmetric
parts, then the solution looks like a three-dimensional monopole
soliton.

We have studied the stability of the obtained 3D analytical
solutions by numerically simulating the evolution of these soliton
solutions in the framework of the original dynamic equation. The
3D vortex soliton without the superimposed parts turns out to be
extremely stable. It moves without distortion and retains its
shape even in the presence of a sufficiently strong initial noise
disturbance. The solitons with parts that are radially symmetric
and/or $z$-antisymmetric are unstable, although at sufficiently
small amplitudes of these superimposed parts, the soliton retains
its shape for a very long time. Due to stability, the predicted 3D
dipole vortices (without the superimposed parts) in the form of a
cyclone-anticyclone pair should apparently exist in astrophysical
objects. For example, the observation of the Mrk 266 galaxy with
two nuclei, rotating in the opposite direction was reported in
Refs.~\cite{Abrahamyan2020,Petrosyan1980}.

\section{ACKNOWLEDGMENTS}

V.M.L. and O.K.C. were supported by the Targeted Complex Program
of the National Academy of Sciences of Ukraine in Plasma Physics.
O.K.C. was also supported by the Thematic Program of the Wolfgang
Pauli Institute 'Models in Plasma, Earth and Space Sciences'.

\section{Appendix}
In problems of nonlinear geophysics, where the nonlinearity is
present in the form of the Jacobian (Poisson brackets), one of the
most reliable numerical methods for representing such a
nonlinearity is the Arakawa scheme \cite{Arakawa1966}. It is based
on the finite difference method. Arbitrary functions $p(x,y,z)$
and $q(x,y,z)$ are represented by its values at the discrete set
of points $x_{i}$, $y_{j}$ and $z_{k}$, We write $p_{i,j,k}$ for
$p(x_{i},y_{j},z_{k})$, and the same for $q$. Since the
nonlinearity in the form of the Poisson bracket ${p,q}$ does not
contain derivatives with respect to $z$, the generalization of the
well-known 2D Arakawa scheme to the 3D case is almost trivial. The
corresponding nonlinearity is written as
\begin{equation}
\{p,q\}=\frac{1}{3}(J^{++}+J^{\times +}+J^{+\times}),
\end{equation}
where
\begin{gather}
J^{++}=\frac{1}{4h_{x}h_{y}}\left[(p_{i+1,j,k}-p_{i-1,j,k})(q_{i,j+1,k}-q_{i,j-1,k})\right.
\nonumber \\ \left.
 -(p_{i,j+1,k}-p_{i,j-1,k})(q_{i+1,j,k}-q_{i-1,j,k})\right],
\\
J^{\times
+}=\frac{1}{4h_{x}h_{y}}\left[q_{i,j+1,k}(p_{i+1,j+1,k}-p_{i-1,j+1,k})
\right. \nonumber \\ \left.
 -q_{i,j-1,k}(p_{i+1,j-1,k}-p_{i-1,j-1,k})
 \right. \nonumber \\ \left.
 -q_{i+1,j,k}(p_{i+1,j+1,k}-p_{i+1,j-1,k})
 \right. \nonumber \\ \left.
 +q_{i-1,j,k}(p_{i-1,j+1,k}-p_{i-1,j-1,k})
\right],
\\
J^{+\times}=\frac{1}{4h_{x}h_{y}}\left[p_{i+1,j,k}(q_{i+1,j+1,k}-q_{i+1,j-1,k})
\right. \nonumber \\ \left.
 -p_{i-1,j,k}(q_{i-1,j+1,k}-q_{i-1,j-1,k})
 \right. \nonumber \\ \left.
 -p_{i,j+1,k}(q_{i+1,j+1,k}-q_{i-1,j+1,k})
 \right. \nonumber \\ \left.
 +p_{i,j-1,k}(q_{i+1,j-1,k}-q_{i-1,j-1,k})
\right],
\end{gather}
where $h_{x}$ and $h_{y}$ are the corresponding grid spacings in
the $x$ and $y$ directions.

\end{document}